\documentclass[12pt]{article}

\usepackage{amssymb,latexsym,amsmath}

\begin{document}

\sloppy
\renewcommand{\theequation}{\arabic{section}.\arabic{equation}}
\thinmuskip = 0.5\thinmuskip    
\medmuskip = 0.5\medmuskip
\thickmuskip = 0.5\thickmuskip
\arraycolsep = 0.3\arraycolsep

\newtheorem{theorem}{Theorem}[section]
\newtheorem{lemma}[theorem]{Lemma}
\newtheorem{prop}[theorem]{Proposition}
\renewcommand{\thetheorem}{\arabic{section}.\arabic{theorem}}

\newcommand{\prf}{\noindent{\bf Proof.}\ }
\def\prfe{\hspace*{\fill} $\Box$

\smallskip \noindent}

\def\be{\begin{equation}}
\def\ee{\end{equation}}
\def\bea{\begin{eqnarray}}
\def\eea{\end{eqnarray}}
\def\beas{\begin{eqnarray*}}
\def\eeas{\end{eqnarray*}}

\newcommand{\R}{\mathbb R} 
\newcommand{\N}{\mathbb N}

\def\supp{\mathrm{supp}\,} 
\def\vol{\mathrm{vol}\,} 
\def\sign{\mathrm{sign}\,}
\def\ekin{E_\mathrm{kin}}
\def\epot{E_\mathrm{pot}}

\def\C{{\cal C}}
\def\H{{\cal H}}
\def\Hc{{{\cal H}_C}}
\def\D{{\cal D}_{f_0}}  
\def\open#1{\setbox0=\hbox{$#1$}
\baselineskip = 0pt
\vbox{\hbox{\hspace*{0.4 \wd0}\tiny $\circ$}\hbox{$#1$}} 
\baselineskip = 11pt\!}
\def\fn{\open{f}}
\def\Rn{\open{R}}
\def\Pn{\open{P}}

\title{Global existence and nonlinear stability for the relativistic
       Vlasov-Poisson system in the gravitational case}
\author{ Mahir Had\v{z}i\'{c}\\
         Division of Applied Mathematics \\
         Brown University, Providence, RI 02912, U.S.A.\\
         and \\
         Gerhard Rein\\
         Mathematisches Institut der
         Universit\"at Bayreuth\\
         D 95440 Bayreuth, Germany}
\maketitle

\begin{abstract}
As is well known from the work of {\sc R.~Glassey}
and {\sc J.~Schaeffer} \cite{GlSch}, the main energy estimates 
which are used in global existence results for the gravitational 
Vlasov-Poisson system do not apply to the relativistic
version of this system, and smooth solutions to the
initial value problem with spherically symmetric initial data of 
negative energy blow up in finite time. 
For similar reasons the variational techniques
by which {\sc Y.~Guo} and {\sc G.~Rein}
obtained nonlinear stability results for
the Vlasov-Poisson system 
\cite{G1,G2,GR1,GR2,R1,R2}
do not apply in the relativistic situation.
In the present paper a direct, non-variational
approach is used to prove nonlinear stability of certain
steady states of the relativistic Vlasov-Poisson system
against spherically symmetric, dynamically accessible
perturbations. The resulting stability estimates
imply that smooth solutions with spherically symmetric initial data
which are sufficiently close to the stable steady states
exist globally in time. 

\end{abstract}

\section{Introduction}
\setcounter{equation}{0}

The topic of the present investigation is the following nonlinear 
system of partial differential equations,
known as the relativistic Vlasov-Poisson system:
\be
\partial_{t}f+\frac{v}{\sqrt{1+|v|^2}}\cdot \nabla_{x}f-
\nabla_{x}U\cdot \nabla_{v}f =0,
\label{vlasov}
\ee
\be
\Delta U = 4\pi \rho,\ \lim_{|x| \to \infty} U(t,x) = 0, \label{poisson}
\ee
\be
\rho(t,x) = \int f(t,x,v)dv \label{rhodef}.  
\ee
Here $t\in\R$ denotes time, $x,v\in \R^3$ denote position and momentum,
and $f=f(t,x,v)\geq 0$ is the time-dependent density
on phase space of a large ensemble of particles which
interact only by the Newtonian gravitational potential
$U=U(t,x)$ which the ensemble creates collectively through
its spatial mass density $\rho=\rho(t,x)$.
Collisions among the particles are assumed to be sufficiently rare to
be neglected. All particles are assumed to be of the same
rest mass, and all physical constants such as the rest mass of a particle,
the gravitational constant, and the
speed of light are normalized to unity. The individual particles
obey the following equations of motion, which form the
characteristic system of the Vlasov equation~(\ref{vlasov}):
\[
\dot x = \frac{v}{\sqrt{1+|v|^2}},\ 
\dot v = - \nabla_{x}U(t,x).
\]
The system is called relativistic because of the relation
between the momentum $v$ of a particle and its velocity $\dot x$.
However, the system as a whole is neither Lorentz nor Galilei
invariant.

In astrophysics galaxies or globular clusters are often modeled 
as large ensembles of particles, i.e., stars, in this way.
But from the point of view of
such applications the non-relativistic Vlasov-Poisson system
where $v/\sqrt{1+|v|^2}$ is replaced by $v$ in the Vlasov
equation is much more important, while a truly relativistic 
formulation is given by the Vlasov-Einstein system \cite{And}.
From a mathematics point of view the non-relativistic Vlasov-Poisson 
system is by now quite well understood. In particular,
the initial value problem has global weak and, for smooth initial data,
global classical solutions, cf.\ \cite{LP,Pf,Sch}.
Moreover, nonlinearly stable steady states
have been established by variational techniques, i.e., 
by minimizing suitably chosen energy-Casimir functionals,
cf.\ \cite{G1,G2,GR1,GR2,LMR,R1,R2}. The stability question
has also received a lot of attention in the astrophysics
literature; we refer to \cite{BT,FP} and the references there.
An extensive review of the mathematical results on global
existence and stability for the non-relativistic system
can be found in \cite{R3}.

For the relativistic version of the system stated above
both the global existence results and the variational 
approach to stability fail, and we briefly explain why.
The total energy
\[
\ekin (f) + \epot (f)
:= \iint \sqrt{1+|v|^2} f(x,v)\,dv\,dx - 
\frac{1}{8 \pi} \int |\nabla U_f(x)|^2 dx
\]
of a state $f$ with induced potential $U_f$
is conserved along solutions,
and so is the mass $||f||_1$ or any other $L^p$ norm $||f||_p$.
Whether one wishes to minimize the energy in a variational stability
analysis or to extract an a-priori bound from the energy
for a global existence result, in both cases one needs
to control the negative potential energy in terms of
a power less than one of the kinetic energy and the other
conserved quantities. But the Hardy-Littlewood-Sobolev inequality
together with standard interpolation arguments imply that
\[
- \epot (f)
\leq
C ||f||_1^{(2-k)/3} ||f||_{1+1/k}^{(k+1)/3} \ekin (f),
\]
where $0\leq k \leq 2$, so no matter which $k$ we choose,
in the context of the above estimate $\epot (f)$ is of the
same order as $\ekin (f)$.
This is the reason why solutions can blow up in finite time,
as shown in \cite{GlSch} for spherically symmetric initial data of negative
energy, and why the variational stability approach
fails. For the non-relativistic system $|v|^2/2$ replaces $\sqrt{1+|v|^2}$
in the kinetic energy, and as a consequence the latter
appears with the exponent $1/2$ in the above estimate.

Recently a non-variational stability approach was introduced
for the Vlasov-Poisson system \cite{GR3}. Its advantage is that
it can deal with steady states which are only local minimizers
of an energy-Casimir functional and not global ones, in 
particular, it requires no lower bound on the energy-Casimir 
functional. For the reason explained above this becomes
essential in the case of the relativistic Vlasov-Poisson
system, and it is the aim of the present investigation to
show that this new method applies to a variety of steady states
of the relativistic Vlasov-Poisson system and provides their
stability against spherically symmetric, dynamically accessible 
perturbations---to remove the symmetry
assumption is an open problem.
In addition we show that the resulting stability estimates
provide bounds on spherically symmetric solutions starting
close to a stable steady state, which are sufficient to prove
global existence. To our knowledge the resulting global existence
results are the first such results for initial data which are
not small.

The paper proceeds as follows. In the next section we state and
explain our main results. Section~3 contains the stability analysis.
In the Newtonian case it has been known for a long time in the
astrophysics literature that the quadratic term in an expansion
of a suitable energy-Casimir functional is indeed positive
definite on the set of linearly dynamically accessible
perturbations. We prove such a result, which can be used to
study linearized stability, for the relativistic Vlasov-Poisson
system. The essential difficulty is then to connect this linear
result to the nonlinear problem. Here our analysis
proceeds essentially as in \cite{GR3}, but we strive for 
greater generality as far as the admissible steady states 
are concerned, because as opposed to the Newtonian case no
other stability results are available here. In Section~4
we briefly show how the stability estimates imply global
existence for the corresponding, perturbed solutions.
In the last section we present a large variety of steady states
to which our method applies. So far only the existence of the
so-called polytropic steady states has been established
for the relativistic Vlasov-Poisson system \cite{Ba}.

\section{Main results}
\setcounter{equation}{0}
Consider a steady state $f_0$ of the relativistic Vlasov-Poisson
system with induced potential $U_0$ 
and spatial density $\rho_0$, where $f_0$ is a function
of the particle energy: 
\be \label{isotropic}
f_0(x,v)=\phi(E)\ \mbox{where}\ E=E(x,v):=\sqrt{1+|v|^2}+U_0(x).
\ee
For a time-independent potential $U_0$ the particle energy $E$ is conserved
along characteristics of the Vlasov equation so that any
function of the particle energy satisfies the Vlasov
equation with potential $U_0$, and it remains to show that 
for a given choice of $\phi$ the
Poisson equation for $U_0$ has a solution, where the spatial density
now becomes a functional of $U_0$. In Section~\ref{stst} we
present a variety of functions $\phi$ for which this approach
gives a compactly supported steady state with finite mass.
Steady states of the form (\ref{isotropic}) are called {\em isotropic},
and it can be shown that they are always spherically symmetric,
i.e.,
\[
f_0(x,v) = f_0(Ax,Av),\ A \in \mathrm{SO}(3),\ x,v \in \R^3.
\]
We want to analyze the stability of such isotropic steady states.
The total energy
\[
\H (f) := \ekin (f) + \epot (f)
= \iint \sqrt{1+|v|^2} f(x,v)\,dv\,dx - 
\frac{1}{8 \pi} \int |\nabla U_f(x)|^2 dx
\]
of a state $f$ is conserved along smooth solutions of the relativistic 
Vlasov-Poisson system. But if we expand $\H$ about 
any state $f_0$ with potential $U_0$ we find that
\[
\H(f) = \H(f_0) + 
\iint \left(\sqrt{1+|v|^2} + U_0 \right)(f-f_0)  \,dv\,dx
- \frac{1}{8 \pi} \int|\nabla U_f-\nabla U_0 |^2 dx,
\]
and the linear part in the expansion does not vanish. Hence we cannot
use the energy as a Lyapunov function in a stability analysis. 
To remedy this situation we observe that 
for any reasonable function $\Phi$ the {\em Casimir functional}
\[
\C(f) := \iint \Phi(f(x,v))\,dv\,dx
\]
is conserved as well. If we expand the energy-Casimir functional 
\[
\Hc := \H + \C
\]
about an isotropic steady state, then with $E$ defined as 
in Eqn.~(\ref{isotropic}),
\beas
\Hc (f) 
&=& 
\Hc (f_0) + 
\iint (E + \Phi'(f_0))\,(f-f_0)  \,dv\,dx \nonumber \\
&& 
{}- \frac{1}{8 \pi} \int|\nabla U_f-\nabla U_0 |^2 dx 
+ \frac{1}{2} \iint \Phi''(f_0) (f-f_0)^2 \,dv\,dx 
+ \ldots .\eeas
We want to choose $\Phi$ is such a way that at least formally
$f_0$ is a critical point of the energy-Casimir functional,
i.e., $\Phi^{\prime }(f_{0})= -E$. In order to make this rigorous
we specify the following assumptions on $\phi$ and $f_0$:

\smallskip

\noindent
{\bf Assumptions on $\phi$ and $f_0$}: (a) $\phi\in C(\R)$, there exists a
cut-off energy $E_0$ such that $\phi(E)=0$ for $E\geq E_0$,
$\lim_{E\to -\infty} \phi(E) = \infty$, and $\phi \in C^2(]-\infty,E_0[)$ with 
\[
\phi'(E)<0\ \mbox{for}\ E<E_0,\ \mbox{and}\
\liminf_{E\to E_0-}\phi'(E)>-\infty. 
\]
(b) $f_0\in C_c(\R^6)$ is compactly supported, satisfies the relation
(\ref{isotropic}), where $U_0$ denotes the
potential induced by $f_0$, and is non-trivial.

\smallskip

\noindent
Since 
\[
\supp f_0 = \{(x,v)\in \R^6 \mid E(x,v) = \sqrt{1+|v|^2} + U_0(x) \leq E_0\}
\]
and since $U_0$ is spherically symmetric and strictly increasing
as a function of $r=|x|$ with $\lim_{r \to \infty} U_0(r) = 0$,
part (b) necessarily implies that $E_0<1$.
In the last section we provide a large class of steady states
which satisfy these assumptions. Examples are 
polytropic steady states
\[
\phi(E)=(E_0-E)_+^k,
\]
with $1\leq k<7/2$,
and the King model
\[
\phi(E)=\left(e^{E_0-E}-1\right)_+;
\]
$(\cdot)_+$ denotes the positive part.

For a function $\phi$ satisfying the assumption above, 
$\phi : ]-\infty ,E_0] \to [0,\infty[$ is continuous
and invertible with continuous inverse
$\phi^{-1} : [0,\infty[ \to ]-\infty ,E_0]$, and
we define
$\Phi: [0,\infty[ \to \R$ by 
\be \label{Phidef}
\Phi(f):= -\int_0^f\phi^{-1}(z)\,dz,\ f\in [0,\infty[.
\ee
In particular, $\Phi\in C^1([0,\infty[)$. 

A crucial step in the stability analysis is to specify the set of admissible
perturbations. From a physics point of view perturbations arise
by some exterior force field acting on the ensemble represented by the
steady state $f_0$. Such a field induces a measure
preserving flow on phase space which redistributes the particles.
We refer to perturbations of the form $f=f_0\circ T$ with 
$T:\R^6 \to \R^6$ a measure preserving $C^1$-diffeomorphism as
{\em dynamically accessible from} $f_0$. For technical reasons we have to
restrict ourselves to spherically symmetric such perturbations. More
precisely, we say that the $C^1$-diffeomorphism $T:\R^6 \to \R^6$
{\em respects spherical symmetry} if for all $x,v \in \R^3$ and all
rotations $A\in \mathrm{SO}(3)$,
\[
T(Ax,Av)=(Ax',Av')\
\mbox{and}\ |x' \times v'| = |x\times v|,\ \mbox{where}\ (x',v')=T(x,v).
\]
Such a redistribution of particles on phase
space would be caused by the action of a field which is spherically symmetric,
and from a physics point of view this restriction is undesirable.
The set of admissible perturbations is defined as
\beas
\D := \Bigl\{ f=f_0 \circ T
&\mid&
T:\R^6 \to \R^6 \ \mbox{is a measure preserving $C^1$-diffeomorphism}\\
&&
\mbox{which respects spherical symmetry}\Bigr\}\,.
\eeas
It is important to note that this set is invariant under
classical solutions of the relativistic Vlasov-Poisson system.
We are going to measure the distance of a state $f\in \D$ 
from the steady state $f_0$ by the quantity
\[
d(f,f_0) := \iint[\Phi(f)-\Phi(f_0)+E (f-f_0)]\,dv\, dx
+\frac{1}{8\pi}\int|\nabla U_f-\nabla U_0|^2\,dx,
\]
which is closely related to the energy-Casimir functional:
\be \label{decrel}
d(f,f_0)= \Hc(f)- \Hc(f_0)+\frac{1}{4\pi}\int|\nabla U_f-\nabla U_0|^2\,dx.
\ee
As we will see in the next section there exists a constant $C>0$ which depends
only on the steady state $f_0$ such that
\be \label{control}
||f - f_0||_2^2 + ||\nabla U_f-\nabla U_0||_2^2 \leq C d(f,f_0),\ f\in \D.
\ee
The major part of the analysis will be concerned with proving
the following result which says---in a precise, quantified manner---that
the steady state is a local minimizer of the energy-Casimir functional
in the set $\D$.
\begin{theorem}\label{th:first}
There exist constants $\delta_0>0$ and $C_0>0$ such that for 
all $f\in \D$ with $d(f,f_0)\leq\delta_0$ the following estimate holds:
\[
\Hc(f)- \Hc(f_0)\geq C_0||\nabla U_f-\nabla U_0||_2^2.
\]
\end{theorem}
This result is proven in the next section. 
Our main result is the following theorem, which is an immediate corollary.
\begin{theorem}\label{th:main}
There exist constants $\delta >0$ and $C>0$ such that 
for any initial datum $\fn \in \D$ with  
\[
d(\fn,f_{0}) < \delta 
\]
the corresponding solution $t\mapsto f(t)$ of the relativistic Vlasov-Poisson
system with $f(0)=\fn$
exists globally in time and satisfies the estimate
\[
d(f(t),f_{0}) \leq C \; d(\fn,f_{0}),\ t\geq 0.
\]
\end{theorem}
\prf 
Let $\delta:= \delta_0 (1+1/(4 \pi C_0))^{-1}$
with $\delta_0$ and $C_0$ from Theorem~\ref{th:first}. Consider a solution
$[0,T[\ni t\mapsto f(t)$ of the relativistic Vlasov-Poisson system with 
$\fn \in \D$ on some maximal interval of existence; as to such a local
existence result we refer to the comments in Section~\ref{sexist}.
Now assume that 
\[
d(\fn,f_{0}) < \delta < \delta_0. 
\]
By continuity we can choose some maximal $t^\ast \in ]0,T]$ such that
\[
d(f(t),f_{0}) < \delta_0,\ t\in [0,t^\ast[. 
\]
Now $f(t) \in \D$ for all $t\in [0,T[$, and hence
Theorem~\ref{th:first}, the relation (\ref{decrel}) of $d$ to the 
energy-Casimir functional, and the fact that the latter is a 
conserved quantity yield the following chain of estimates for 
$t\in[0,t^\ast[$:
\beas
d(f(t),f_0)
&=&
\Hc(f(t)) - \Hc(f_0) + \frac{1}{4\pi} ||\nabla U_{f(t)}-\nabla U_0||_2^2\\
&\leq&
\Hc(f(t)) - \Hc(f_0) + \frac{1}{4 \pi C_0} \left(\Hc(f(t)) - \Hc(f_0)\right)\\
&=&
\left(1+\frac{1}{4 \pi C_0}\right) \, \left(\Hc(f(0)) - \Hc(f_0)\right)
\leq
\left(1+\frac{1}{4 \pi C_0}\right)\,d(\fn,f_0) < \delta_0.
\eeas
This implies that $t^\ast = T$. By Proposition~\ref{exist}, $T=\infty$, 
and Theorem~\ref{th:main} is established.
\prfe

\smallskip

\noindent
{\bf Remark.}
Theorems~\ref{th:first} and \ref{th:main} are analogous to the
corresponding results obtained for the non-relativistic case in \cite{GR3},
but there are two differences. Firstly, the analysis in \cite{GR3}
is restricted to the King model, the reason being that stability for other
models had already been obtained via variational methods.
Since these methods do not work in the relativistic context,
as was explained in the introduction, we keep the admissible steady states
more general here, but the same generality would have been possible
in the non-relativistic case. Secondly, for the non-relativistic case
global existence of classical solutions is known for general data,
independently of any stability analysis. In the relativistic case this 
is not so. Theorem~\ref{th:main} includes a new global existence result
for the relativistic system for data which are not subject to a size
restriction. This should be compared with the results in \cite{GlSch},
in particular, with the fact that spherically symmetric solutions
with $\H (\fn)<0$ blow up in finite time. 
In particular, Theorem~\ref{th:main} 
answers an open problem which was formulated in \cite[Final remarks]{Ba}.

\section{Proof of Theorem~\ref{th:first}}
\setcounter{equation}{0}
The essence of the proof of Theorem~\ref{th:first}
is the analysis of the quadratic term 
\be \label{h2}
D^2\Hc (f_0)[g] :=  \frac{1}{2}\iint_{\{f_{0}>0\}}
\Phi''(f_0) g^2\, dv\, dx 
-\frac{1}{8\pi}\int |\nabla U_{g}|^2 dx  
\ee
which arises in the expansion of the energy-Casimir functional $\Hc$.
If Theorem~\ref{th:first} were false, then a tangent direction
$g$ to the set $\D$ of dynamically accessible states would exist
on which this quadratic part is negative, cf.\ Lemma~\ref{lm:findg}.
On the other hand, if we define the Poisson bracket
of two functions $f,h : \R^6 \to \R$ by
\be \label{poissonbr}
\{ f,h\} := \nabla_x f \cdot \nabla_v h - \nabla_v f \cdot \nabla_x h,
\ee
then  for {\em linearized, dynamically accessible states}
$g = \{f_0,h\}$ 
one can show that the quadratic term is indeed positive definite,
cf.\ Lemma~\ref{lm:kandrup}; we do not go into the symplectic
dynamics details behind this terminology and construction. Note however
that at least formally, states of this bracket form arise as tangent
vectors to the manifold $\D$ at the point $f_0$, and that the set
of these states is invariant under the linearized
system. The analogue of Lemma~\ref{lm:kandrup} for the non-relativistic
case is well known in the astrophysics literature, where it has been used
to prove linearized stability \cite{KS,Aly1,SDLP}, 
but for the relativistic case the
result seems to be new. To conclude the proof of Theorem~\ref{th:first}
we then show that the function $g$ with $D^2\Hc (f_0)[g] \leq 0$ obtained 
in Lemma~\ref{lm:findg} can be written in the form  $g = \{f_0,h\}$,
which contradicts Lemma~\ref{lm:kandrup}.
At some points the arguments are similar to the non-relativistic
case considered in \cite{GR3}. But since we consider a
different system and also a general class
of admissible steady states and not just one example, the 
analysis is technically more difficult, and we prefer to give a
self-contained proof for the present situation.

At several points we have to exploit the spherical symmetry
assumption which is part of our definition of the set $\D$ of
dynamically accessible data.  To do so we on occasion
use coordinates which are adapted to the spherical symmetry:
\be \label{sscoord}
r:= |x|,\ w:= \frac{x\cdot v}{r},\ L:= |x\times v|^2;
\ee
$w$ is the radial component of momentum and $L$ is the modulus of
angular momentum squared.
It is easy to see that for any spherically symmetric function
$f$ by abuse of notation, $f(x,v) = f(r,w,L)$.

We first collect some properties
of the function $\Phi$ defined in Eqn.~(\ref{Phidef}). 
\begin{lemma}\label{lm:useful}
\begin{itemize}
\item[\rm{(a)}]
$\Phi \in C^1([0,\infty[) \cap C^3(]0,\infty[)$, $\Phi(f) \geq -E_0 f$ for
$f\geq 0$, and for $f>0$,
\[
\Phi'(f)=-\phi^{-1}(f),\
\Phi''(f)=- \frac{1}{\phi'(\phi^{-1}(f))},\
\Phi'''(f)=\frac{\phi''(\phi^{-1}(f))}{(\phi'(\phi^{-1}(f)))^3}.
\]
\item[\rm{(b)}]
There exists $C>0$ such that for all functions $f\in \D$,
\[
\iint\left[\Phi(f)-\Phi(f_0) + E (f-f_0)\right]\, dv\, dx 
\geq C \iint |f-f_0|^2\, dv\, dx,
\]
in particular, Eqn.~(\ref{control}) holds.
\end{itemize}
\end{lemma}
\prf
The formulas for the derivatives are obvious,
and the lower bound for $\Phi$ follows from the fact that 
$\phi^{-1} \leq E_0$.
The assumptions on $\phi'$ imply that
\[
C := \frac{1}{2} \inf \{ \Phi''(f) \mid 0<f\leq ||f_0||_\infty +1\} >0.
\] 
For $f\in \D$,
\[
E (f-f_0) = - \Phi'(f_0) (f-f_0) \ \mbox{on}\ \{f_0 > 0\}
\]
while
\[
E (f-f_0) = E  f \geq E_0 f = - \Phi'(0) f = - \Phi'(f_0) (f-f_0)
\ \mbox{on}\ \{f_0 = 0\}.
\]
Hence by Taylor's Theorem
\beas
\Phi(f)-\Phi(f_0) + E (f-f_0)
&\geq&
\Phi(f)-\Phi(f_0) -\Phi'(f_0)(f-f_0)\\
&=&
\lim_{\epsilon\to 0+} 
\left[\Phi(f+\epsilon)-\Phi(f_0+\epsilon) -\Phi'(f_0+\epsilon)(f-f_0)\right] \\
&=&
\lim_{\epsilon\to 0+}\frac{1}{2}\Phi''(\xi_\epsilon) |f-f_0|^2 \geq C \,|f-f_0|^2,
\eeas
since $0< \xi_\epsilon \leq ||f_0||_\infty +1$ as $\epsilon \to 0$, 
and the proof is complete.
\prfe

Assuming that Theorem~\ref{th:first} were false we construct a
state $g$ on which the quadratic term $D^2 \Hc (f_0)$ is negative,
more precisely:
\begin{lemma}\label{lm:findg}
Assume that Theorem~\ref{th:first} were false. Then there 
exists a function $g\in L^2(\mathbb R^6)$ which is spherically symmetric, 
supported in $\supp f_0$, even in $v$, i.e., $g(x,-v)=g(x,v)$, 
and such that
\be \label{eq:first}
\frac{1}{8\pi}||\nabla U_g||_2^2=1,
\ee
\be \label{eq:second}
D^2 \Hc (f_0)[g]=\frac{1}{2}\iint_{\{f_0>0\}}
\Phi''(f_0)g^2\,dv\,dx - 1\leq 0,
\ee
and for all functions $G=G(f,L)\in C^2 ([0,\infty[^2)$
with $G(0,L) = \partial_f G(0,L) = 0$ for $L\geq 0$ and 
$\partial_f^2 G$ bounded,  
\be\label{eq:third}
\iint\partial_f G (f_0,L)g\,dv\,dx =0 .
\ee
\end{lemma}
{\bf Remark.} Eqn.~(\ref{eq:third}) expresses the fact that $g$
is a direction tangent to $\D$ at $f_0$. If $\tau \mapsto f(\tau)$ is
a curve on $\D$ with $f(0)=f_0$ and $\frac{d}{d\tau} f(0) = g$
then by definition of $\D$, $\iint G(f(\tau),L) = \iint G(f_0,L)$
so that the formal derivative of the left hand side at $\tau=0$
vanishes, which is Eqn.~(\ref{eq:third}).

\smallskip

\noindent
{\bf Proof of Lemma~\ref{lm:findg}.}
Since we assume that Theorem~\ref{th:first} is false, 
there exists a sequence $(f_n)\subset \D$ such that
for all $n\in \N$,
\[
d(f_n,f_0)<\frac{1}{n},
\]
but
\begin{equation}\label{eq:2}
 \Hc(f_n)- \Hc(f_0)<\frac{1}{8\pi n}||\nabla U_{f_n}-\nabla U_0||_2^2.
\end{equation}
If we let
\begin{equation}\label{eq:3.6}
f_n=f_0+\sigma_n g_n \ \mbox{with}\
\frac{1}{8\pi}||\nabla U_{g_n}||_2^2=1,
\end{equation}
i.e.,
\[
\sigma_n:=\frac{1}{\sqrt{8\pi}} ||\nabla U_{f_n}-\nabla U_0||_2,\
g_n:=\frac{1}{\sigma_n}(f_n-f_0),
\]
then in particular,
\be \label{sigmazero}
\sigma_n^2\leq d(f_n,f_0)<\frac{1}{n}.
\ee
{\em A weak limit $g$ of a subsequence of $(g_n)$.}\\
Using the definition of $d$ and Eqns.~(\ref{eq:2}), (\ref{eq:3.6}), and
(\ref{decrel}) we find that 
\beas
&&
\frac{1}{\sigma _{n}^2}\iint
[\Phi (f_n)-\Phi (f_{0})+ E\,\sigma _{n}g_{n}]
\,dv\,dx  - 1 \nonumber \\
&&
\qquad\qquad = \frac{1}{\sigma _{n}^2}\left(d(f_n,f_0)
-\frac{1}{4\pi} ||\nabla U_{f_{n}}-\nabla U_0||_2^2 \right) \nonumber\\ 
&&
\qquad\qquad = \frac{1}{\sigma _{n}^2} \left(\Hc(f_{n})-\Hc(f_{0})\right)
< \frac{1}{\sigma _{n}^2} 
\frac{1}{8\pi\, n } ||\nabla U_{f_{n}}-\nabla U_0||_2^2 
= \frac{1}{n}.\quad 
\eeas
By Lemma~\ref{lm:useful} this implies that
\be \label{gnbound}
1+\frac{1}{n}
>
\frac{1}{\sigma_n^2}
\iint [\Phi (f_n)-\Phi (f_{0})+ E\,\sigma _{n}g_{n}]
\,dv\,dx\\
\geq 
C\iint|g_n|^2\,dv\,dx,
\ee
and hence the sequence $(g_n)$ is bounded in $L^2(\R^6)$.
We extract a subsequence, again denoted by $(g_n)$, such that
\[
g_n \rightharpoonup g\ \mbox{weakly in}\ L^2(\R^6).
\]
Since the functions $g_n$
are spherically symmetric so is $g$,
and we need to show that it is supported on $\supp f_0$.
The term in brackets in Eqn.~(\ref{gnbound}) is non-negative,
and on the set $\{E > E_0\}$ the steady
state distribution $f_0$ and hence also $\Phi(f_0)$ vanish,
while $\Phi(f_n)\geq -E_0 f_n = -E_0 \sigma_n g_n$. Hence
\[
2
> \frac{1}{\sigma _{n}^2}
\iint_{\{E > E_0\}} 
[\Phi (f_n)-\Phi (f_{0})+ E\,\sigma _{n}g_{n}]\,dv\,dx 
\geq
\frac{1}{\sigma _{n}}
\iint_{\{E > E_0\}} (E-E_0)\, g_n\,dv\,dx,
\]
and by (\ref{sigmazero}),
\be \label{zerooutside1}
\iint_{\{E > E_0\}} (E-E_0)\, g_n\,dv\,dx \leq 2 \sigma_n \to 0,\
n\to \infty.
\ee
For any fixed $E_0< E_1 < 1$, 
\be \label{zerooutside2}
\iint_{\{E > E_1\}} g_n\,dv\,dx 
\leq 
\iint_{\{E > E_0\}} \frac{E-E_0}{E_1-E_0} \, g_n\,dv\,dx \to 0;
\ee
notice that $g_n \geq 0$ on the set $\{E > E_0\}$. 
Since (\ref{zerooutside2}) holds
for any $E_0< E_1 < 1$, we conclude that $g$ 
is supported on $\supp f_0 = \{E\leq E_0\}$. 
Moreover, since for $E_0< E_1 < 1$ fixed the set $\{E \leq E_1\}$ 
is compact, the $L^2$-bound on $(g_n)$ and Eqn.~(\ref{zerooutside2}) 
imply that  $(g_n)$ is bounded in $L^1(\R^6)$ as well.

\smallskip

\noindent
{\em Proof of (\ref{eq:first}).}\\
First we note that
$U_0$ is spherically symmetric and radially increasing, 
so in particular, $U_0(0) \leq U_0 \leq 0$.
Hence 
\beas
\iint\sqrt{1+|v|^2}|g_n|\,dv\,dx
&\leq& 
\iint E\,|g_n|\,dv\,dx -U_0(0) \iint |g_n|\,dv\,dx\\
&\leq&
E_0 \iint |g_n|\,dv\,dx 
+ \iint_{\{E > E_0\}} (E-E_0) |g_n|\,dv\,dx\\
&&
{}- U_0(0) \iint |g_n|\,dv\,dx,
\eeas
so that together with (\ref{zerooutside1}) and the $L^1$-bound on $(g_n)$
the kinetic energy is bounded along $(g_n)$.
By well known interpolation arguments
\cite[Ch.~1, Lemma~5.1]{R3}
the sequence of induced spatial densities $(\rho_{g_n})$ 
is bounded in $L^{5/4} (\R^3)$, 
so without loss of generality this sequence converges 
weakly in $L^{5/4}(\R^3)$.
We fix $E_0<E_1<1$ and $R_1>0$
such that $1+ U_0(R_1)=E_1$, which implies that $E(x,v)\geq E_1$
for $|x|\geq R_1$. Then by (\ref{zerooutside2}),
\[
\int_{\{|x| > R_1\}} |\rho_{g_n}|\, dx
\leq
\iint_{\{E > E_1\}} g_n\,dv\,dx
\to 0,
\]
i.e., the sequence $(\rho_{g_n})$ remains concentrated, and hence
\[
\nabla U_{g_n} \to \nabla U_g \ \mbox{strongly in}\ L^2(\R^3),
\]
cf.\ \cite[Ch.~2, Lemma~3.2]{R3}. 
Passing to the limit in Eqn.~(\ref{eq:3.6}) proves that
$g$ satisfies the Eqn.~(\ref{eq:first}).

\smallskip

\noindent
{\em Proof of (\ref{eq:second}).}\\
We first claim that there exists a sequence of sets 
$K_j\subset K_{j+1}\subset\ldots \subset \supp f_0$ such that 
for a subsequence which we again denote by $(g_n)$,
\[
\vol (\supp f_0 \setminus K_j)<\frac{1}{j}\ 
\mbox{and}\ \lim_{n\to\infty}
\sigma_n g_n=0\ \mbox{uniformly on}\ K_j,\ j\in \N.
\]
To see this we note that 
$||\sigma _{n}g_{n}||_2 \leq C\sigma _{n} \to 0$ 
so that a subsequence converges to zero pointwise a.~e.. Then we use
Egorov's theorem on the set $\supp f_0$ which has finite measure. 
In the following arguments we need to stay away from the
boundary of the latter set, and hence we define
\[
S_m:=\left\{(x,v)\in\R^6 \mid E(x,v)\leq E_0-1/m\right\},\ m\in \N.
\]
Clearly, $\delta_m:=\inf_{S_m} f_0 >0$. Hence
for all sufficiently large $n$,
\[
\delta_m/2 \leq f_0+\sigma_ng_n 
\leq ||f_0||_\infty +1\ \mbox{on}\ S_m \cap K_j.
\]
On the set $S_m \cap K_j$, 
\[
\Phi(f_n)-\Phi(f_0)+E (f_n-f_0) =
\frac{1}{2}\Phi''(f_0)(\sigma_n g_n)^2 
+
\frac{1}{6}\Phi'''(f_0+\xi\sigma_ng_n)(\sigma_ng_n)^3,
\]
with $0\leq \xi\leq 1$; recall that $\Phi\in C^3(]0,\infty[)$.
In particular, 
\[
|\Phi'''(f_0+\xi\sigma_ng_n)| \leq 
\sup\left\{|\Phi'''(z)| \mid
\delta_m/2 \leq z \leq ||f_0||_\infty +1\right\} =: C_m < \infty.
\]
Using (\ref{gnbound}) we find
that for all sufficiently large $n$,
\begin{eqnarray*}
\frac{1}{2} \iint_{S_m \cap K_j} \Phi''(f_0)|g_n|^2\,dv\,dx
&=&
\frac{1}{\sigma_n^2}\iint_{S_m \cap K_j}
[\Phi(f_n)-\Phi(f_0)+E (f_n-f_0)]\,dv\,dx\\
&&
{} -\frac{1}{6 \sigma_n^2} \iint_{S_m \cap K_j}
\Phi'''(f_0+\xi\sigma_n g_n)(\sigma_ng_n)^3\,dv\,dx\\
&<&
1+\frac{1}{n}+C_m
\sup_{K_j}|\sigma_ng_n|\iint|g_n|^2\,dv\, dx.
\end{eqnarray*}
Now
$ g_{n}\rightharpoonup g$ weakly in $L^2(\R^6)$ 
and $\sigma_n g_{n}\to 0$ uniformly on $K_j$.
Taking the limit $n\to\infty$ implies that for all $j,m \in \N$, 
\[
\frac{1}{2} \iint_{S_m \cap K_j} \Phi''(f_0) g^2\,dv\,dx \leq 1;
\]
the latter expression is $L^2$-weakly lower semicontinuous in $g$.
By the boundedness of $\Phi''(f_0)$ on $S_m$ and
the choice of $K_j$ we find with $j\to\infty$ that
\[
\frac{1}{2} \iint_{S_m} \Phi''(f_0) g^2\,dv\,dx \leq 1,
\]
and with $m\to \infty$
the monotone convergence theorem implies Eqn.~(\ref{eq:second}).

\smallskip

\noindent
{\em Proof of (\ref{eq:third}).}\\ 
Let $G=G(f,L)$ be a function as specified in Eqn.~(\ref{eq:third}).
By Taylor expansion with respect to the first argument,
\[
G(f_n,L)-G(f_{0},L)
=\partial _{f}G (f_{0},L)\,\sigma _{n}g_{n}+
\frac{1}{2}\partial_{f}^2 G(f_{0}+\tau\sigma_{n}g_{n},L)\,
(\sigma _{n}g_{n})^2
\]
for some $\tau\in[0,1]$. For $f\in \D$, 
$
\iint G(f,L) =\iint G(f_{0},L),
$
and hence
\[
\iint
\partial_{f} G(f_{0},L)\,g_{n}\,dv\,dx 
=
-\frac{1}{2}\sigma_{n}
\iint\partial_f^2 G (f_{0}+\tau \sigma_{n}g_{n},L)\,g_n^2\,dv\,dx
\to 0;
\]
note that $\partial_f^2 G$ is bounded,
$(g_n)$ is bounded in $L^2(\R^6)$, and $\sigma _{n}\to 0$. 
On the other hand $\partial_{f} G(f_{0},L)$ is supported
on the compact set $\supp f_0$ and hence bounded.
Since $g_{n}\rightharpoonup g$ weakly in $L^2(\R^6)$, 
Eqn.~(\ref{eq:third}) follows as $n \to \infty$.

\smallskip

\noindent
{\em Conclusion of the proof of Lemma~\ref{lm:findg}.}\\
The function $g$ constructed above has all the required properties, 
except that it need not be even in $v$. 
However, if we decompose it into its even and odd parts with respect to 
$v$, $g=g_\mathrm{even}+g_\mathrm{odd}$, then
$g_\mathrm{even}$ satisfies  
(\ref{eq:first}), (\ref{eq:second}), (\ref{eq:third}) 
as well. Since 
$\rho_{g}=\rho_{g_\mathrm{even}}$
we have $\nabla U_{g_\mathrm{even}}=\nabla U_{g}$, and (\ref{eq:first})
remains valid. 
Since $\Phi''(f_0)\geq 0$ is even in $v$, 
\beas
1
&\geq&
\frac{1}{2} \iint_{\{f_0>0\}} 
\Phi''(f_0)\, (g_\mathrm{even}+g_\mathrm{odd})^2
=
\frac{1}{2} \iint_{\{f_0>0\}} 
\Phi''(f_0)\, \left((g_\mathrm{even})^2+(g_\mathrm{odd})^2\right) \\
&\geq&
\frac{1}{2}\iint_{\{f_0>0\}}
\Phi''(f_0)\, (g_\mathrm{even})^2,
\eeas
i.e., (\ref{eq:second}) remains valid. Finally, for $G$ as in
(\ref{eq:third}),
$\partial_f G(f_0,L)$ is even in $v$ so that
the odd part of $g$ drops out of Eqn.~(\ref{eq:third}),
and the proof of Lemma~\ref{lm:findg} is complete.
\prfe

As was said above for states of the form $g=\{f_0,h\}$
the quadratic term in the expansion of $\Hc$ is positive
definite; the Poisson bracket was defined in Eqn.~(\ref{poissonbr}).
It turns out that it will make some arguments technically
easier later on, if we prove this fact for states of the
form $g=\{E,h\}$. 
\begin{lemma}\label{lm:kandrup}
Let $h\in C_c^{\infty}(\mathbb R^6)$ be spherically symmetric 
with $\supp h \subset \{f_0>0\}$ and such that $h(x,-v)=-h(x,v)$. 
Then the following inequality holds:
\begin{eqnarray*}
&&
D^2\Hc(f_0)[\{E,h\}]\\
&&
\qquad \qquad \geq
-\frac{1}{2} \iint \frac{1}{\phi'(E)}
\left[|x\cdot v|^2 \left|\left\{E,\frac{h}{x\cdot v}\right\}\right|^2
+\frac{U_0'h^2}{r(1+|v|^2)^{3/2}} \right]\,dv\,dx.
\end{eqnarray*}
\end{lemma}
\prf
Let
\[
U_h(x):=\iint\frac{\{E,h\}}{|x-y|}\,dv\,dy
\]
denote the potential induced by $-\{E,h\}$.
Using the definition of the Poisson bracket,
\[
\int\{E,h\}dv=\nabla_x\cdot\int\frac{v}{\sqrt{1+|v|^2}} h(x,v)\,dv.
\]
Since both $h$ and $U_h$ are spherically symmetric,
\begin{eqnarray*}
U_h'(r)=4\pi\int\frac{w}{\sqrt{1+|v|^2}} h(x,v)\,dv.
\end{eqnarray*}
By the Cauchy-Schwarz inequality,
\[
\frac{1}{8\pi}\int|\nabla U_h|^2\,dx
\leq
2\pi\int
\left[\int\frac{-w^2}{\sqrt{1+|v|^2}}\phi'(E)\,dv\right]
\left[\int\frac{1}{\sqrt{1+|v|^2}} \frac{-h^2}{\phi'(E)}
dv\right]\,dx.
\]
Since
\[
\frac{w^2}{\sqrt{1+|v|^2}}\phi'(E)=w\frac{d}{dw}\phi
\left(\sqrt{1+w^2+L/r^2}+U_0(r)\right)
=w\frac{d}{dw}\phi(E),
\]
an integration by parts with respect to $w$ yields
\[
\int\frac{-w^2}{\sqrt{1+|v|^2}}\phi'(E)\,dv=\rho_0(r):= \rho_{f_0}(r).
\]
Hence
\[
D^2\Hc(f_0)[\{E,h\}]
\geq
-\frac{1}{2}\iint \frac{1}{\phi'(E)}
\left[|\{E,h\}|^2
-4\pi\frac{\rho_0(r)}{\sqrt{1+|v|^2}}h^2\right]\,dv\,dx.
\]
Since $h$ is odd in $v$ and thus in $w$ the function 
\[
\mu(r,w,L):=\frac{1}{rw}h(r,w,L)
\]
is smooth away from $r=0$. Using the identity $h=rw\mu$ 
a straight forward computation shows that 
\[
\big|\{E,h\}\big|^2=(rw)^2\big|\{E,\mu\}\big|^2+\{E,\mu^2rw\{E,rw\}\}-
\mu^2rw\{E,\{E,rw\}\}.
\]
The first term is as claimed in the lemma. The second term 
leads to
$\{\psi(E),\mu^2rw\{E,rw\}\}$ with $\psi$ a primitive of
$1/\phi'$,
and the integral of this expression with respect to $x$ and $v$ 
vanishes via an integration by parts; 
if we cut a small ball of radius $\epsilon$
about $x=0$ from the $x$-integral
then the surface integral appearing after the integration by parts
with respect to $x$ vanishes for $\epsilon \to 0$ since $r \mu^2 \leq C/r$. 
Finally, another lengthy, but straight forward computation
using the fact that $\Delta U_0 = 4 \pi \rho_0$ implies that
\[
-\mu^2rw\{E,\{E,rw\}\}=\frac{1}{1+|v|^2}\frac{h^2}{\sqrt{1+|v|^2}}
\frac{U_0'(r)}{r}+4\pi\rho_0(r)\frac{h^2}{\sqrt{1+|v|^2}}.
\]
If we substitute this into the above estimate for $D^2\Hc$
the proof of Lemma~\ref{lm:kandrup} is complete.
\prfe

In order to derive a contradiction between Lemma~\ref{lm:kandrup}
and Lemma~\ref{lm:findg} we have to show that the function $g$
provided by the latter can be written in the form $\{E,h\}$.
Taking the opposite sign is of course equivalent but more convenient
in the sequel.

\smallskip

\noindent
{\em Solving $g=\{-E,h\}$.}\\
Let $g$ be the function obtained in Lemma~\ref{lm:findg}. In the 
spherical variables introduced in (\ref{sscoord}) the equation to be 
solved for $h$ takes the form
\begin{equation}\label{eq:findh}
\frac{w}{\sqrt{1+w^2+L/r^2}}\partial_r h+
\left(\frac{L}{r^3\sqrt{1+w^2+L/r^2}}-U_0'(r)\right)\partial_w h=g.
\end{equation}
The characteristic system of this equation reads
\begin{equation}\label{eq:characteristics}
\dot{r}=\frac{w}{\sqrt{1+w^2+L/r^2}},\quad
\dot{w}=\frac{L}{r^3\sqrt{1+w^2+L/r^2}}-U_0'(r).
\end{equation}
In order to analyze it, we introduce for fixed $L>0$ the function 
\[
\Psi_L(r)=U_0(r)+\sqrt{1+L/r^2}.
\]
In the spherical variables the particle energy takes the form
\be \label{ssen}
E=E(x,v) = E(r,w,L):=\sqrt{1+w^2+L/r^2}+U_0(r).
\ee
It is conserved along solutions of the characteristic 
system (\ref{eq:characteristics}), while $L$ only plays the role of
a parameter. For given values of $E$ and $L$ we want to
identify the $r$-interval in which a corresponding characteristic can range.
By spherical symmetry,
\[
U_0'(r)=\frac{m_0(r)}{r^2},\ \mbox{where}\
m_0(r):=4\pi\int_0^r\rho_0(s)s^2\,ds,\ r>0.
\] 
Since the steady state is non-trivial and $U_0$ increasing, $U_0(0)<E_0-1$, 
and by (\ref{isotropic}),
$\rho_0(0)>0$. Next,
\[
\Psi_L'(r)=0\ \Leftrightarrow\ m_0(r)-\frac{L}{\sqrt{L+r^2}}=0,
\]
and since the left hand side of the latter equation is strictly 
increasing for $L>0$ with limit $-\sqrt{L}$ for $r\to 0$ and 
$||f_0||_1 >0$ for $r\to \infty$, 
there exists a unique radius $r_L>0$ such that 
\[
\Psi_L'(r_L)=0,\ \Psi_L'(r)<0\ \mbox{for}\ r<r_L,\ 
\Psi_L'(r)>0\ \mbox{for}\ r>r_L.
\]
Since $\lim_{r\to 0} \Psi_L(r) =\infty$ and 
$\lim_{r\to \infty} \Psi_L(r) =1$ 
we conclude that for any $L>0$ and $\Psi_L(r_L)<E<1$ there exist 
unique radii $0<r_-(E,L)<r_L<r_+(E,L)<\infty$ such that 
\[
\Psi_L(r_\pm(E,L))=E,\ \mbox{and}\ 
\Psi_L(r)<E\;\Leftrightarrow\; r_-(E,L)<r<r_+(E,L).
\]
For later use we note that by the 
implicit function theorem the mapping 
$]0,\infty[\ni L\mapsto r_L$ 
is continuously differentiable,
since $\frac{d}{dr}(m_0(r)-L/\sqrt{L+r^2})>0$. 
The same is true for the mapping
$(E,L)\mapsto r_\pm(E,L)$ 
on the set $\{(E,L)\in\mathbb R\times]0,\infty[ \mid \Psi_L(r_L)<E<1\}$,
since $\Psi_L'(r)\neq0$ for $r\neq r_L$.
We also note that 
\begin{equation}\label{eq:sec.der}
\Psi_L''(r_L)=4\pi\rho_0(r_L)+\frac{L}{r_L(L+r_L^2)^{3/2}}>0.
\end{equation} 
If $h$ is to solve Eqn.~(\ref{eq:findh}), then for any solution
$\tau \mapsto (r(\tau),w(\tau),L)$ of the characteristic system
(\ref{eq:characteristics}),
\[
\frac{d}{d\tau}h(r(\tau),w(\tau),L)=g(r(\tau),w(\tau),L).
\]
As long as $w\neq 0$ we can rewrite this in terms of the variable $r$:
\[
\frac{d}{dr}h(r,w(r),L)=\frac{\sqrt{1+w^2(r)+L/r^2}}{w(r)} g(r,w(r),L).
\]
Since $E$ and $L$ are constant along characteristics we
express $w$ in terms of these conserved quantities via the relation
(\ref{ssen}). More precisely, with
\be \label{wrel}
w(r,E,L):= \sqrt{(E-U_0(r))^2-L/r^2-1},\ q(r,E,L) := \frac{E-U_0(r)}{w(r,E,L)}
\ee
we have $w(r)= \sign w(r)\, w(r,E,L)$ and 
\[
\frac{d}{dr}h(r,w(r),L)= \sign w(r) q(r,E,L) g(r,w(r,E,L),L).
\]
Outside $\supp f_0$ we set $h=0$. 
For $(r,w,L)\in\supp f_0$ with $L>0$ we define $h$ as follows. 
We let $E:=E(r,w,L)$ so that $r_-(E,L) \leq r\leq r_+(E,L)$, 
and
\begin{equation}\label{eq:h}
h(r,w,L):=\sign w \int_{r_-(E,L)}^r g(s,w(s,E,L),L)\, q(s,E,L)\,ds.
\end{equation}
In order for this definition to
be consistent we show that the integral above vanishes if $r=r_+(E,L)$
so that  $h(r,0,L)=0$. To this end let $G$ be as 
specified in Eqn.~(\ref{eq:third}) so that
$\iint \partial_f G(\phi(E),L) g =0$. We want to use the change of variables
$(x,v) \mapsto (r,w,L) \mapsto (r,E,L)$. Since $g$ is even in $v$ 
and hence in $w$ we can extend this integral only over $\{w>0\}$, and
\be \label{xvtorel}
dv\,dx = 8\pi^2\,dr\,dw\,dL=8\pi^2 q(r,E,L)\,dr\,dE\,dL.
\ee
Thus
\[
\iint_M \int_{r_-(E,L)}^{r_+(E,L)} g(r,w(r,E,L),L) 
\, q(r,E,L)\,dr\,
\partial_{f} G(\phi(E),L)\,dE\,dL = 0,
\]
where $M:=\{(E,L)(x,v) \mid f_0(x,v)>0\}$. The class of test functions 
$\partial_f G(\phi(E),L)$ is large enough to conclude that for almost 
all $E$ and $L$ the integral with respect to $r$ vanishes as desired.

As defined above the function $h$ is not sufficiently regular to
apply Lemma~\ref{lm:kandrup} to it.  
Fubini's Theorem shows that $h$ is measurable,
because by the change of variables formula the function
\beas
(s,r,E,L) 
&\mapsto& 
\mathbf{1}_{[\Psi_L(r_L),E_0]}(E)\mathbf{1}_{[r_-(E,L),r_+(E,L)]}(s)
\mathbf{1}_{[0,r]}(s)\\
&&
\qquad \qquad \qquad \qquad \qquad
g(s,w(s,E,L),L) \, q(s,E,L) 
\eeas
is integrable; $r\leq \max\{|x| \mid (x,v) \in \supp f_0\}$,
and $\mathbf{1}_M$ denotes the indicator function of the set $M$.

\smallskip

\noindent
{\em The cut-off $h$.}\\
We need to regularize $h$, and as a first step we
study a cut-off version of $h$. To do so we need an 
auxiliary result.
\begin{lemma} \label{line}
For every $m\in \N$ there exists a constant $C_m>0$ such that for $L\geq 1/m$ 
and $\Psi_L(r_L)<E\leq E_0$,
\[
\int_{r_-(E,L)}^{r_+(E,L)} q(r,E,L)\,dr<C_m.
\]
\end{lemma}
\prf
We first claim that for all $m\in\N$ there exists a constant $\eta_m>0$ 
such that for all $L\geq 1/m,\,\Psi_L(r_L)<E\leq E_0$, and 
$r\in[r_-(E,L),r_+(E,L)]$,
\begin{equation}\label{eq:cleverestimate}
\frac{|\Psi_L'(r)|}{\sqrt{\Psi_L(r)-\Psi_L(r_L)}}\geq\eta_m.
\end{equation}
To see this, let $m\in \N$ and $L,E,r$ be as specified. 
Then
\[
E_0\geq E\geq\Psi_L(r)=U_0(r)+\sqrt{1+L/r^2}\geq U_0(0)+\sqrt{1+1/(m r^2)},
\]
and hence $r \geq \left( m\,((E_0-U_0(0))^2-1)\right)^{-1/2}$. 
Let $R:=\max\{|x| \mid (x,v)\in\supp f_0\}$. 
Then
\[
L\leq r^2\left((E_0-U_0(r))^2-1\right)\leq R^2
\left((E_0-U_0(0))^2-1\right).
\]
If the above claim were false, there would
exist a sequence $(r_n,L_n)\to(\bar{r},\bar{L})$ in the set 
$[\left(m((E_0-U_0(0))^2-1)\right)^{-1/2},R]\times[1/m,
R^2\left((E_0-U_0(0))^2-1\right)]$, 
such that 
\begin{equation} \label{contra}
\lim_{n\to\infty}
\frac{\Psi_{L_n}'(r_n)}{\sqrt{\Psi_{L_n}(r_n)-\Psi_{L_n}(r_{L_n})}}=0.
\end{equation}
We can assume that $\bar{r}=r_{\bar{L}}$ since otherwise 
$\Psi_{\bar{L}}(\bar{r})>\Psi_{\bar{L}}(r_{\bar{L}})$ 
and $\Psi_{\bar{L}}'(\bar{r})=0$ 
which is a contradiction to the uniqueness of the minimizer 
$r_{\bar{L}}$ of $\Psi_{\bar{L}}$. By Taylor expansion at $r=r_{L_n}$ we 
find intermediate values $\theta_n$ and $\tau_n$ between $r_n$ and $r_{L_n}$ 
such that
\[
\frac{|\Psi_{L_n}'(r_n)|}{\sqrt{\Psi_{L_n}(r_n)-\Psi_{L_n}(r_{L_n})}}=
\frac{\big|\Psi_{L_n}''(\theta_n)(r_n-r_{L_n})\big|}
{\sqrt{\frac{1}{2}\Psi_{L_n}''(\tau_n)(r_n-r_{L_n})^2}}
\to\sqrt{2|\Psi_{\bar{L}}''(r_{\bar{L}})}|,\ n\to\infty,
\]
so that (\ref{contra}) contradicts (\ref{eq:sec.der}), 
and (\ref{eq:cleverestimate}) is established.

We split the integral under investigation into two parts,
\[
\int_{r_-(E,L)}^{r_+(E,L)} q(r,E,L)\,dr
=\int_{r_-(E,L)}^{r_L}\dots+\int_{r_L}^{r_+(E,L)}\dots=:I_1+I_2.
\] 
Using  the definitions of $w(r,E,L)$ and $\Psi_L$ and the fact that 
$E-U_0(r)\geq\Psi_L(r)-U_0(r)\geq1$ we find that 
\begin{eqnarray*}
q(r,E,L)=\frac{E-U_0(r)}{w(r,E,L)}
&=&
\frac{E-U_0(r)}{\sqrt{(E-U_0(r))^2-(\Psi_L(r)-U_0(r))^2}}\\
&=&
\frac{E-U_0(r)}{\sqrt{(E-\Psi_L(r))(E+\Psi_L(r)-2U_0(r))}}\\
&\leq&
\frac{E-U_0(0)}{\sqrt{2(E-\Psi_L(r))}}\leq\frac{C}{\sqrt{E-\Psi_L(r)}}
\end{eqnarray*}
for some constant $C$ which does not depend on $r,E$, or $L$. 
In the integral $I_1$ we change variables $u=\sqrt{\Psi_L(r)-\Psi_L (r_L)}$ 
so that $\frac{du}{dr}=\frac{1}{2u}\Psi_L'(r)<0$ on $[r_-(E,L),r_L[$. 
By~(\ref{eq:cleverestimate}) and the previous estimate,
\begin{eqnarray*}
I_1
&\leq&
C \int_{\sqrt{E-\Psi_L(r_L)}}^0
\frac{1}{\sqrt{E-\Psi_L(r_L)-u^2}}\frac{dr}{du}du \\
&\leq& 
\frac{1}{\eta_m}\int_0^{\sqrt{E-\Psi_L(r_L)}}
\frac{du}{\sqrt{E-\Psi_L(r_L)-u^2}}
=
\frac{1}{\eta_m}\int_0^1\frac{ds}{\sqrt{1-s^2}}<\infty.
\end{eqnarray*}
We argue similarly 
for $I_2$, and the proof of the lemma is complete.
\prfe

For $m\in \N$ let 
\[
\Omega_m := \left\{ (x,v) \in \R^6 \mid 
E(x,v) \leq E_0 -\frac{1}{m},\ L(x,v) \geq  \frac{1}{m} \right\}.
\]
With the help of the previous lemma we can prove that 
$\mathbf{1}_{\Omega_m}h$ is square integrable and solves the 
equation $\{-E,\mathbf{1}_{\Omega_m} h\}=\mathbf{1}_{\Omega_m} g$ 
in the sense of distributions.
\begin{lemma}\label{lm:distributions}
For $m\in\mathbb N$ large, $\mathbf{1}_{\Omega_m}h\in L^2(\R^6)$, 
and for any spherically symmetric test function 
$\psi=\psi(r,w,L)\in C^1([0,\infty[\times\R\times[0,\infty[)$,
\[
\iint \{-E,\psi\} \mathbf{1}_{\Omega_m} h \,dv\,dx =
-\iint \psi \mathbf{1}_{\Omega_m} g \,dv\,dx .
\]
\end{lemma}
\prf
In order to prove that $\mathbf{1}_{\Omega _{m}} h\in L^2(\R^6)$ we observe 
that the integrand is even in $v$ so that we can apply
the change of variables (\ref{xvtorel}):
\[
\iint \mathbf{1}_{\Omega _{m}} h^2 dv\,dx
= 8 \pi^2 \iint_{S_m} 
\int_{r_{-}(E,L)}^{r_{+}(E,L)}h^2(r,w(r,E,L),L)
\, q(r,E,L)\,dr\,dE\,dL,
\]
where 
\[
S_m:= 
\left\{(E,L)=(E,L)(x,v)
\mid (x,v)\in \Omega_m\right\}.
\]
Let $(E,L) \in S_m$. Writing 
$w(r) =w(r,E,L)$,
$q(r)=q(r,E,L)$
and
$r_\pm = r_\pm(E,L)$
for brevity the definition
(\ref{eq:h}) of $h$, the Cauchy-Schwarz inequality, and Lemma~\ref{line}
imply that
\begin{eqnarray*}
\int_{r_{-}}^{r_{+}}h^2(r,w(r),L) q(r)\,dr
&=&
\int_{r_{-}}^{r_{+}}
\left[\int_{r_{-}}^{r}
g(s,w(s),L) q(s)\,ds \right]^2 q(r)\,dr \\
&\leq&
\left(\int_{r_{-}}^{r_{+}} q(r)\, dr\right)^2
\int_{r_{-}}^{r_{+}}
g^2(r,w(r),L) q(r) dr\\
&\leq&
C_m^2
\int_{r_{-}}^{r_{+}}
g^2(r,w(r),L) q(r)\, dr.
\end{eqnarray*}
A further integration with respect to $E$ and $L$ and the change
of variables  $(r,E,L)\mapsto (x,v)$ shows that 
$||\mathbf{1}_{\Omega _{m}} h||_2$ is bounded in terms of $C_m$ and
$||g||_2$.

Let $\psi$ be a test function as specified in the lemma. 
As above, we parameterize the solutions of
(\ref{eq:characteristics}) by  $r$, distinguishing between $w>0$ 
and $w<0$. Then it is easy to show that 
\[
\{-E,\psi\}=\textrm{sign}\,w\frac{1}{q(r)}\frac{d}{dr}[\psi(r,w(r),L)].
\] 
Using the same change of variables as before and 
recalling Eqn.~(\ref{eq:h}) we find that
\begin{eqnarray*}
&&
\iint\{-E,\psi\} \mathbf{1}_{\Omega_m} h\,dv\,dx
=\int_{\{w>0\}}\dots+\int_{\{w<0\}}\dots\\
&&
\qquad 
=8 \pi^2 \iint_{S_m}\int_{r_-}^{r_+}\frac{1}{q(r)}\frac{d}{dr}
[\psi(r,w(r),L)]\,h(r,w(r),L) \, q(r)\,dr\,dE\,dL\\
&&
\quad \qquad
- 8 \pi^2 \iint_{S_m}\int_{r_-}^{r_+}\frac{1}{q(r)}\frac{d}{dr}
[\psi(r,-w(r),L)]\,h(r,-w(r),L)\,q(r)\,dr\,dE\,dL\\
&&
\qquad 
= - 8 \pi^2 \iint_{S_m}\int_{r_-}^{r_+}\psi(r,w(r),L)\,
\, g(r,w(r),L)\,q(r)\,dr\,dE\,dL\\
&&
\quad \qquad 
-8 \pi^2 \iint_{S_m}\int_{r_-}^{r_+}\psi(r,-w(r),L)
\,g(r,-w(r),L)\,q(r)\,dr\,dE\,dL\\
&&
\qquad
= -\iint\psi \mathbf{1}_{\Omega_m} g\,dv\,dx;
\end{eqnarray*}
note that $h(r,\pm w(r),L)=0$ for $r=r_{\pm}(E,L)$ 
which together with the definition of $h$ is used in the integration 
by parts above, and also that $g$ is even both in $v$ and $w$. 
The proof of Lemma~\ref{lm:distributions} is complete.
\prfe

\smallskip

\noindent
{\em Regularization of $h$.}\\
The function $\mathbf{1}_{\Omega_m} h$ is not smooth, so
in order to apply Lemma~\ref{lm:kandrup} we smooth 
it. For $m\in\N$ fixed the 
function $\mathbf{1}_{\Omega_m} h$ is, as a function of $r,w,L$, 
supported in a cube of the form
\[
Q := [R_0,R_1]\times[-W_0,W_0]\times[L_0,L_1],
\]
with $0<R_0<R_1$, $W_0>0$, and $0<L_0<L_1$. Let 
$\zeta\in C_c^{\infty}(\R^3)$ be even in all three variables, i.e., 
$\zeta(z_1,z_2,z_3)=\zeta(|z_1|,|z_2|,|z_3|)$, $\zeta\geq0,\,\int\zeta=1$, 
and let $\zeta_n(z):=n^3\zeta(n z)$ for $n\in\N$. For $n$ 
sufficiently large we define
\[
h_n(r,w,L):=\int_0^{\infty}\int_{-\infty}^{\infty}\int_0^{\infty}
(\mathbf{1}_{\Omega_m} h)(\bar{r},\bar{w},\bar{L})
\zeta_n(r-\bar{r},w-\bar{w},L-\bar{L})\,d\bar{L}\,d\bar{w}\,d\bar{r}.
\]
Clearly, $h_n\in C_c^{\infty}(]0,\infty[\times\R\times]0,\infty[)$, 
and without loss of generality we can assume that
$\supp h_n\subset Q \cap \{E<E_0\}$. 
Since $h$ is odd in $w$, so is $h_n$, and clearly 
$h_n\to \mathbf{1}_{\Omega_m} h$ in $L^1\cap L^2$. The crucial step 
is to show that 
\begin{equation}\label{eq:crucial}
\lim_{n\to\infty}\{-E,h_n\}=\mathbf{1}_{\Omega_m} g\,\,\,\,\textrm{in}\,\,L^2.
\end{equation}
We fix $(r,w,L)$, write 
$p:= \sqrt{1+w^2+L/r^2}$, 
$\bar p:= \sqrt{1+\bar w^2+\bar L/\bar r^2}$,
and 
$\int:=\int_0^{\infty}\int_{-\infty}^{\infty}\int_0^{\infty}$ 
for brevity, and split the convolution integral as follows:
\begin{eqnarray*}
&&\{-E,h_n\}
=
\frac{w}{p} \partial_r h_n +
\left(\frac{L}{r^3 p}-U_0'(r)\right)\partial_w h_n\\
&&
\qquad =
\int (\mathbf{1}_{\Omega_m}h)(\bar{r},\bar{w},\bar{L})
\left[\left(\frac{\bar{w}}{\bar p}-
\frac{w}{p}\right)\partial_{\bar{r}}
+\left(\frac{\bar{L}}{\bar{r}^3 \bar p}
-\frac{L}{r^3 p}-U_0'(\bar{r})+U_0'(r)\right)\partial_{\bar{w}}\right]\\
&&\qquad\qquad\qquad\qquad\qquad\qquad\qquad\qquad\qquad\quad
\zeta_n(r-\bar{r},w-\bar{w},L-\bar{L})
\,d\bar{r}\,d\bar{w}\,d\bar{L}\\
&&
\qquad \quad
{} - \int (\mathbf{1}_{\Omega_m}h)(\bar{r},\bar{w},\bar{L})
\left[\frac{\bar{w}}{\bar p}
\partial_{\bar{r}}+\left(\frac{\bar{L}}{\bar{r}^3 \bar p}
-U_0'(\bar{r})\right)\partial_{\bar{w}}\right]\\
&&\qquad\qquad\qquad\qquad\qquad\qquad\qquad\qquad\qquad\quad
\zeta_n (r-\bar{r},w-\bar{w},L-\bar{L})\,d\bar{r}\,d\bar{w}\,d\bar{L}\\
&&
\qquad
=:
 J_1+J_2.
\end{eqnarray*}
According to Lemma~\ref{lm:distributions}, 
\begin{eqnarray*}
J_2
&=&
-\int \mathbf{1}_{\Omega_m} h\, \{-E,\zeta_n(r-\cdot,w-\cdot,L-\cdot)\}\\
&=&
\int (\mathbf{1}_{\Omega_m} g) (\bar{r},\bar{w},\bar{L})\,
\zeta_n(r-\bar{r},w-\bar{w},L-\bar{L})\,d\bar{r}\,d\bar{w}\,d\bar{L}
\to \mathbf{1}_{\Omega_m} g\ \mbox{in}\ L^2.
\end{eqnarray*}
We show that $J_1$ converges to $0$ as $n\to\infty$. To this end
we introduce new variables $\tilde{r}=n(r-\bar{r})$, $\tilde{w}=n(w-\bar{w})$, 
$\tilde{L}=n(L-\bar{L})$. Then $\bar{r}=r-\tilde{r}/n$, 
$\bar{w}=w-\tilde{w}/n$, $\bar{L}=L-\tilde{L}/n$, 
$d\bar{r}\,d\bar{w}\,d\bar{L}=n^{-3}d\tilde{r}\,d\tilde{w}\,d\tilde{L}$, and
$\partial_r\zeta_n(r-\bar{r},w-\bar{w},L-\bar{L})=n^4\partial_{\tilde{r}}
\zeta(\tilde{r},\tilde{w},\tilde{L})$.
Now we note that 
\begin{eqnarray*}
&&
\int(\mathbf{1}_{\Omega_m} h)(\bar{r},\bar{w},\bar{L})
\left(\frac{\bar{w}}{\bar p}-\frac{w}{p}\right)\partial_{\bar{r}}
\zeta_n(r-\bar{r},w-\bar{w},L-\bar{L})\,d\bar{r}\,d\bar{w}\,d\bar{L}\\
&&
\qquad\qquad\qquad
=n\int(\mathbf{1}_{\Omega_m} h)(\bar{r},\bar{w},\bar{L})
\left(\frac{\bar{w}}{\bar p}-\frac{w}{p}\right)\partial_{\tilde{r}}
\zeta_n(\tilde{r},\tilde{w},\tilde{L})\,d\tilde{r}\,d\tilde{w}\,d\tilde{L}.
\end{eqnarray*}
The analogous identity holds for the second part of $J_1$, and we find that
\begin{eqnarray*}
n\left(\frac{\bar{w}}{\bar p}-\frac{w}{p}\right)
&=&
n \frac{\bar{w}-w}{\bar p}
+nw\left(\frac{1}{\bar p}-\frac{1}{p}\right)\\
&=&
- \frac{\tilde{w}}{p}+nw\frac{w^2-\bar{w}^2+L/r^2-
\bar{L}/\bar{r}^2}{\bar p p (\bar p + p)}\\
&\to&
-\frac{\tilde{w}}{p}+
\frac{2\tilde{w}w^2+ w\tilde{L}/r^2- 2\tilde{r}wL/r^3}{2 p^3},\\
\end{eqnarray*}
since
\[
\frac{L}{r^2}-\frac{\bar{L}}{\bar{r}^2}=\frac{L-\bar{L}}
{\bar{r}^2}+L\frac{(\bar{r}-r)(\bar{r}+r)}{r^2\bar{r}^2}.
\]
We easily see that
\[
n(U'_0(r)-U'_0(\bar{r}))\to U_0''(r)\tilde{r}.
\]
Furthermore, 
\beas
n\,\left(\frac{\bar{L}}{\bar{r}^3\bar p}-\frac{L}{r^3 p}\right)
&=&
n\,\left(\frac{\bar{L}/\bar{r}^3 - L/r^3}{\bar p}+
\frac{L}{r^3}\left(\frac{1}{\bar p}-\frac{1}{p}\right)\right)\\
&\to&
\frac{1}{p}\left(\frac{3 L\tilde{r}}{r^4}-\frac{\tilde{L}}{r^3}\right)+
\frac{L}{r^3}\frac{2\tilde{w}w+
\tilde{L}/r^2 - 2\tilde{r}L/r^3}{2 p^3}.
\eeas
Thus most of the terms above integrate to $0$ in the 
limit of the expression 
$J_1$, leaving us in the end with
\[
J_1 \to
(\mathbf{1}_{\Omega_m} h)(r,w,L)\frac{Lw}{r^3 p^3}
\int\big(-\tilde{r}\partial_{\tilde{r}}+\tilde{w}\partial_{\tilde{w}}\big)
\zeta(\tilde{r},\tilde{w},\tilde{L})\,d\tilde{r}\,d\tilde{w}\,d\tilde{L}=0,
\]
where the limit is in $L^2$, and $0$ arises in the last equation 
from an integration by parts with respect to $\tilde{r}$ and 
$\tilde{w}$ and an exact cancellation. This completes the 
proof of~(\ref{eq:crucial}). 

\smallskip

\noindent
{\em The contradiction to Lemma~\ref{lm:kandrup}.}\\
The functions $h_n$ have all the properties required in 
Lemma~\ref{lm:kandrup}. 
By Eqn.~(\ref{eq:crucial}), 
$\lim_{n\to\infty}\{-E,h_n\} = \mathbf{1}_{\Omega_m}g$ in $L^2$, 
and since $h_n$ and $\mathbf{1}_{\Omega_m} g$ are supported 
in a common compact set, 
$\nabla U_{\{-E,h_n\}}\to\nabla U_{\mathbf{1}_{\Omega_m}g}$ 
in $L^2$ as $n\to\infty$. This implies that
\[
D^2\Hc(f_0)[\{-E,h_n\}] \to D^2\Hc(f_0)[\mathbf{1}_{\Omega_m} g],
\ n\to \infty.
\]
Since $\lim_{m\to\infty} \mathbf{1}_{\Omega_m}g =g$ in $L^1\cap L^2(\R^6)$, 
there exists $m_0$ sufficiently large such that 
$\mathbf{1}_{\Omega_{m_0}}g\neq 0$, and by Lemma~\ref{lm:distributions}, 
$\mathbf{1}_{\Omega_{m_0}} h \neq 0$. For all $m\geq m_0$, 
$\Omega_{m_0}\subset\Omega_m$ 
and thus $\mathbf{1}_{\Omega_{m_0}} h^2 \leq \mathbf{1}_{\Omega_{m}}h^2$. 
Since $\phi'(E)<0$ and  
$U_0'(r)>0$ 
we have for all $m\geq m_0$,
\[
D^2\Hc(f_0)[\mathbf{1}_{\Omega_m} g]
\geq -\frac{1}{2} \iint
\frac{1}{\phi'(E)} \frac{1}{r} U_0'
\mathbf{1}_{\Omega_{m_0}} h^2 \left(1+|v|^2\right)^{-3/2}\,dv\,dx >0.
\]
Letting $m\to\infty$ we get a contradiction to Lemma~\ref{lm:findg},
and the proof of Theorem~\ref{th:first} is complete.

\section{Global existence of the perturbed solutions} \label{sexist}
\setcounter{equation}{0}
Let $f_0$ be a steady state as specified in Section~2,
and choose $\delta >0$ according to Theorem~\ref{th:main}.
For solutions with initial data $\fn \in \D$ satisfying
the estimate $d(\fn,f_0) < \delta$ 
the potential energy and therefore also the kinetic energy
remain bounded. In the present section we show that
this implies that such solutions are global in time.
\begin{prop} \label{exist}
Let $\fn \in \D$. Then there exists a unique,
smooth solution $f$ of the relativistic Vlasov-Poisson
system with $f(0)=\fn$ on a maximal time interval $[0,T[$
with $T>0$.
If 
\[
\sup_{0\leq t< T} \ekin(f(t)) < \infty,
\]
then this solution is global, i.e., $T=\infty$.
\end{prop}

\prf
It is well known that a compactly supported, non-negative initial 
datum $\fn \in C^1_c(\R^6)$ launches a unique
$C^1$ solution on some time interval $[0,T[$ which we choose maximal.
A detailed proof for the non-relativistic case is given in
\cite{R3}, and the proof carries over to the relativistic situation.
In addition, the solution is known to be global if
\[
\sup \{|v| \mid (x,v) \in \supp f(t),\ 0\leq t<T\} < \infty.
\]
However, there is a technical problem here. The steady state $f_0$
and therefore also data taken from $\D$ need not be continuously
differentiable on $\R^6$, they need only be continuous on $\R^6$
and continuously differentiable on the set where they are
strictly positive. But due to the spherical symmetry
of the functions in $\D$ these regularity properties
propagate to the solutions, since 
\[
\nabla U(t,x) = \frac{4\pi}{r^2} \int_0^r \rho(t,s) s^2 ds \frac{x}{r},\
r=|x|,
\] 
is continuously differentiable if $f$ and therefore $\rho$
are continuous. In particular, the characteristic flow
of the Vlasov equation is well-defined and continuously
differentiable and $f$ is constant along these characteristics.
To turn these arguments into a rigorous proof would require
going through the local existence proof for the present situation,
which would be lengthy but straight forward.

Let us now assume a bound on the kinetic energy of a maximal,
spherically symmetric
solution $f$ on some time interval $[0,T[$. We want to deduce
a bound on the quantity
\[
P(t):= \sup \{ |v| \mid (x,v) \in \supp f(s),\ 0\leq s \leq t \},
\]
which by the continuation criterion recalled above implies
global existence. The arguments which follow are adapted
from \cite{GlSch}, but we prefer to give a self-contained
presentation. First we note that a standard interpolation
estimate implies that $||\rho(t)||_{4/3}$ is bounded
on $[0,T[$, cf.~\cite[Lemma~5.1]{R3}. This in turn implies that
for all $0\leq s \leq t < T$,
\[
||\nabla U(s)||_\infty \leq C ||\rho(s)||_\infty^{5/9}
\leq C P^{5/3} (t),
\]
cf.~\cite[Lemma~P1]{R3};
the constant $C$ depends only on the initial datum $\fn$ and 
on the bound on the kinetic energy. Due to the above formula for the field
in the spherically symmetric case,
\be \label{uprest}
|U'(s,r)| < c \min\{ r^{-2},P^{5/3} (t)\} \leq c (r+a)^{-2},\ r\geq 0,
\ee
where we abbreviate $a:=P^{-5/6} (t)$; the constant $c>0$ is now kept
fixed. 

For $L\geq 0$ we define the function
\[
\xi_L(u):=\sqrt{1+u^2} - \frac{c u}{\sqrt{L}+ a u},\ u \geq 0.
\]
It is easy to check that $\xi_L \in C^2([0,\infty[)$
with $\xi_L'' >0$. Moreover, there exists a unique $u_L >0$
such that $\xi_L(u_L)=1$ and $\xi_L$ is strictly increasing
on the interval $[u_L,\infty[$. Clearly,
\[
\xi_L(u) \geq \xi_0(u)  = \sqrt{1+u^2} - \frac{c}{a}
\]
for all $u\geq 0$ and $L\geq 0$, and this implies that
\[
u_L \leq u_0 = \sqrt{(1+c/a)^2-1}.
\]
Consider now a characteristic 
$[0,T[ \ni s \mapsto (x(s),v(s))$ in the support of $f$. 
With
\[
r(s):=|x(s)|,\ u(s):=|v(s)|,\ w(s):=x(s)\cdot v(s)/r(s),\
L(s):= |x(s)\times v(s)|^2
\]
it follows that $L(s)=L$ is constant and
\[
\frac{d}{ds} u^2(s) = -2 U'(s,r(s))\, w(s).
\]
Let us first consider the case that $L>0$, fix some time
$t\in ]0,T[$, and assume that $w(t)<0$. We choose
$0\leq t_0 < t$ minimal with the property that
$w(s) < 0$ for $s\in ]t_0,t]$. Then
$u(t) \geq u(t_0)$, and
\beas
&&
\frac{d}{ds} \left(\sqrt{1+u^2(s)} -
c \int_{r(s)}^\infty \frac{dr}{(r+a)^2}\right) \\
&&
\qquad \qquad\qquad\qquad\qquad {}=
\left(\frac{c}{ (r(s)+a)^2} - U'(s,r(s))\right)\, \frac{w(s)}{\sqrt{1+u^2(s)}} < 0
\eeas
by the estimate (\ref{uprest}). Since $u^2(t)= w^2(t) + L/r^2(t)$
we have the estimate $r(t) > \sqrt{L}/u(t)$, and hence
\[
\sqrt{1+u^2(t_0)} -
c \int_{r(t_0)}^\infty \frac{dr}{(r+a)^2} > 
\sqrt{1+u^2(t)} -
c \int_{\sqrt{L}/u(t)}^\infty \frac{dr}{(r+a)^2} = \xi_L(u(t)).
\]
If $w(t_0)=0$ then $u(t_0)=\sqrt{L}/r(t_0)$ and hence
\[
\xi_L(u(t)) < \xi_L(u(t_0)),
\]
which by choice of $u_L$ and the fact that $u(t) \geq u(t_0)$
implies that $u(t) \leq u_L \leq u_0$.
If instead $w(t_0)<0$ then $t_0=0$ and $\xi_L(u(t)) \leq \sqrt{1+P^2(0)}$.
Hence in the case $L>0$ and $w(t)<0$ we have shown that
\be \label{uest}
\sqrt{1+u^2(t)} \leq  \sqrt{1+P^2(0)} + c/a;
\ee 
notice the formula for $u_0$ stated above. 

Next we consider the case
that $L>0$ and $w(t)\geq 0$. Again, we choose $0\leq t_0 \leq t$
minimal such that $w(s)\geq 0$ for $s \in [t_0,t]$.
Then $u(t)\leq u(t_0)$. If $t_0=0$ then $u(t) \leq P(0)$.
If $t_0>0$ there must exist a sequence $0<t_i \nearrow t_0$
such that $w(t_i) <0$, which by the case we already dealt with
implies that
\[
\sqrt{1+u^2(t_i)} \leq  \sqrt{1+P^2(0)} + c/a,
\] 
and by continuity, this estimate must also hold with $t_0$ instead
of $t_i$. Hence the estimate (\ref{uest}) holds
for all characteristics with $L>0$, but since the initial
data of a characteristic in $\supp f$ with $L=0$ can be
approximated as closely as we wish by data with $L>0$
and since characteristics depend continuously on their
initial data the estimate (\ref{uest}) remains true if $L=0$.
Hence
\[
\sqrt{1+P^2(t)} \leq  \sqrt{1+P^2(0)} + c/a
= \sqrt{1+P^2(0)} + c\, P^{5/6} (t),\ 0\leq t < T.
\]
This implies that $P$ is bounded on $[0,T[$ by some constant
which depends on the initial datum $\fn$ and the bound
for the kinetic energy, and the proof is complete.
\prfe
\section{Stationary solutions} \label{stst}
\setcounter{equation}{0}
As we explained at the beginning of Section~2
steady states of the relativistic Vlasov-Poisson system
can be obtained by making the ansatz (\ref{isotropic})
for some function $\phi=\phi(E)$. This ansatz 
solves the stationary Vlasov equation with potential $U_0$
in the sense that $f_0$ is constant along characteristics.
By substituting it into the definition of the spatial density 
the latter becomes a functional of the potential $U_0$:
\begin{equation}\label{eq:rho1}
\rho_0 (x)=g_\phi(U_0(x)),
\end{equation}
where
\be \label{gphidef}
g_\phi (u) 
:=4 \pi \int^{\infty}_{u+1}
\phi(E)(E-u)\left[(E-u)^2-1\right]^{1/2}\,dE,\ u\in \R.
\ee
Hence a steady state of the relativistic Vlasov-Poisson system is obtained,
if for a given choice of $\phi$ it can be shown that the semilinear Poisson
equation
\[
\Delta U_0 = 4 \pi g_\phi(U_0)
\]
has a solution on $\R^3$
such that $f_0 := \phi\circ E$ has finite mass and
compact support. Since in this section we consider only
stationary solutions, we drop the subscript $0$ and denote the steady
state by $f$ and its potential by $U$.

One can show that any solution obtained by the above 
approach must a-posteriori be spherically symmetric, so we only look for 
solutions $U=U (r),\ r\geq 0,$ of the equation
\be \label{semilinp}
U'(r) = \frac{4 \pi}{r^2}\int_0^r g_\phi(U(s))\, s^2\, ds,\ r>0.
\ee
In~\cite{Ba}, {\sc Batt} showed that the polytropic ansatz
\[
\phi(E)=(E_0-E)_+^k,
\]
with
$k \in ]0,2[$ leads to steady states with compact support and finite mass.
Below we present two approaches which can be used to prove
the existence of steady states for more general functions $\phi$. 
First however, we show that  $\phi$ must vanish for large 
values of $E$ if the resulting steady state is to have finite mass.
\begin{prop}
Let $\phi:\R\to[0,\infty[$ be measurable and let $(f,U)$ be a stationary
solution of the relativistic Vlasov-Poisson system in the sense
that $f(x,v)=\phi(E)$ and $U\in C^1([0,\infty[)$ solves (\ref{semilinp})
with $g_\phi$ given by (\ref{gphidef}). 
Let $\iint f\, dv\,dx <\infty$. 
Then $U_{\infty}:= \lim_{r\to\infty}U(r)<\infty$ and 
$\phi(E)=0$ a.e. for $E>U_{\infty}+1$.
\end{prop}
\prf
Let $M:=\iint f\, dv\,dx$. Since by (\ref{semilinp}),
$0\leq U'(r)\leq M/r^2$, $U$ is increasing with a finite
limit $U_\infty$ for $r\to\infty$, and since $g_\phi$ is decreasing,
\[
M =
4\pi \int_0^\infty g_\phi(U(r))\, r^2\,dr 
\geq 4\pi \int_0^\infty g_\phi(U_\infty)\, r^2\,dr,
\]
which means that $g_\phi(U_\infty)=0$. This is only true if
$\phi(E)=0$ for almost all $E>U_\infty +1$. 
\prfe
In what follows we are only interested in steady states with finite mass 
so we always assume such a cut-off energy $E_0$.

Under suitable assumptions on $\phi$
Eqn.~(\ref{semilinp}) has a unique solution on $[0,\infty[$,
if we prescribe $U(0)$. 
\begin{prop} \label{globalexss}
Let $\phi:\R\to [0,\infty[$ have the following properties:
$\phi$ is measurable,
there exists a cut-off energy $E_0\in \R$ such that
$\phi(E)=0$ for $E>E_0$, and there exists $k>-1$ such that
for every $E_1 < E_0$ there exists a constant $C\geq 0$
such that
\[
\phi(E) \leq C (E_0-E)^k,\ E\in [E_1,E_0[.
\]
Then for every $u_0\in\R$ there exists 
a unique solution $U\in C^1([0,\infty[)$ of (\ref{semilinp}) with
$U(0)=u_0$.
\end{prop}
\prf
One can show that under the given assumptions on $\phi$
the function $g_\phi$ defined in Eqn.~(\ref{gphidef}) is continuously
differentiable.
The existence and uniqueness of $U$ on some interval $[0,\delta]$
then follows by a contraction argument. Since $U$ is increasing,
either $U \leq E_0-1$ on its maximal existence interval, and 
the solution is global, or there exists some $R\geq 0$
such that $U(r)>E_0-1$ for $r\geq R$, 
which implies that $g_\phi(U(r))=0$ for $r\geq R$, 
and again $U$ is global.
\prfe

\smallskip

\noindent
{\bf Remark.}
The solution obtained above will in general not satisfy
the boundary condition $U_\infty =0$ which we required
in previous sections, and which is part of our formulation
of the relativistic Vlasov-Poisson system.
However, the solution has a finite limit $U_\infty$, and by 
subtracting this limit from $U$ and redefining the cut-off
energy $E_0$ accordingly we obtain a steady state with the
right boundary condition. It is clear that $E_0$ and $U(0)$
cannot be freely prescribed under the given boundary condition
for $U$ at infinity.

\smallskip

We now proceed to our first result which guarantees the existence
of compactly supported steady states with finite mass.
We follow an approach which was used for the Vlasov-Poisson
and Vlasov-Einstein systems in \cite{RR}.
\begin{prop} \label{finradss}
Let $\phi$ be as in Proposition~\ref{globalexss}, and assume in addition
that
\[
\phi(E)=c(E_0-E)^k + O\left((E_0-E)^{k+\delta}\right)\ \mbox{as}\ E\to E_0-, 
\]
for some $-1/2<k< 3/2$, $c>0$, and $\delta>0$. 
Let $(f,U)$ be an induced steady state of the relativistic Vlasov-Poisson system, i.e., 
$f(x,v)=\phi(E)$, and $U\in C^1([0,\infty[)$ satisfies (\ref{semilinp}). 
Then this
steady state has compact support and finite mass.
\end{prop}
It is remarkable that up to technical assumptions only
the behavior of $\phi$ at the cut-off energy $E_0$ needs
to be specified. In particular, there are
plenty of functions $\phi$ which satisfy the assumptions
of this theorem and of the stability result in Section~2
as well. Beside the polytropic states with $1\leq k<3/2$ a
notable example which frequently appears in the astrophysics
literature is a King type model
\[
f(x,v)=\phi(E)=\left(e^{E_0-E}-1\right)_+.
\]

The basic set-up of the proof of Proposition~\ref{finradss}
is as follows. If $U(0) \geq E_0 - 1$ then this
relation holds also for $r>0$, and the steady state is trivial. Hence we
consider a solution $U$ such that $U(0) < E_0 -1$, and we define 
$[0,R[$ as the maximal interval on which $U<E_0-1$
so that $\supp \rho = [0,R]$.
If $R=\infty$ then 
$U_\infty =\lim_{r\to \infty} U(r) \leq E_0-1$ exists by monotonicity. 
Now assume that $U_\infty < E_0 -1$. 
Then $\rho(r) \geq 4\pi g_\phi(U_\infty)>0$ which implies that 
$U'(r) \geq C r$ with $C>0$ which upon integration gives the
contradiction that $U_\infty = \infty$.  Hence
\be \label{setup}
U(r) < E_0 -1 \ \mbox{on}\ [0,R[,\ \mbox{and}\ 
\lim_{r\to R-} U(r) =  E_0 -1;
\ee
if $R<\infty$ the limit assertion is true as well.
In order to show that (\ref{setup}) indeed implies that $R<\infty$
we use a special case of a result due to {\sc Makino}
\cite{mak}; a proof can also be found in \cite[Lemma 3.2]{RR}. 
\begin{lemma}\label{lm:makino}
Let $x,y\in C^1(]0,R[)$ be such that $x,y>0$ and
\beas
r x'
&=&
\alpha(r)y-x+x^2\\
r y'
&=&
y\left(a -\beta(r)x\right)
\eeas
on $]0,R[$, where $a>0$ is a constant, $\alpha, \beta\in C(]0,R[)$, 
$\inf_{r\in]0,R[}\alpha(r)>0$, and $\lim_{r\to R-}\beta(r)\in]0,a[$.
Then $R<\infty$.
\end{lemma} 
{\bf Proof of Proposition~\ref{finradss}}.
We first  introduce the radial pressure
\[
p(x)=p(r) := \int \left(\frac{x\cdot v}{r}\right)^2\, f(x,v)\, dv.
\]
Like the spatial density $\rho$ the pressure becomes a functional of $U$
via Eqn.~(\ref{isotropic}):
\[
p(r) = h_\phi(U(r))
\]
where
\be \label{hphidef}
h_\phi (u) :=
\frac{4 \pi}{3} \int^{\infty}_{u+1}\phi(E)(E-u)
\left[(E-u)^2-1\right]^{3/2}\,dE,\ u\in \R.
\ee
On the interval $]0,R[$ we define
\[
\eta(r) := E_0-U(r)-1,\ m(r):= 4\pi \int_0^rs^2\rho(s)\,ds
\]
and
\[
x(r) :=
\frac{m(r)}{r\eta(r)},\
y(r)
:=
4\pi r^2\frac{\rho^2(r)}{p(r)}.
\]
A short computation shows that the latter functions
satisfy a system of ODEs of the form stated in Lemma~\ref{lm:makino}
with $a:=2$ and 
\beas
\alpha 
&:=&
\frac{p}{\rho \eta} =\frac{h(\eta)}{g (\eta)\,\eta},\\
\beta
&:=&
\left(\frac{rp'}{p} - \frac{2 r \rho'}{\rho}\right)\,
\frac{r \eta}{m} =
2 \eta \frac{g'(\eta)}{g(\eta)} - \eta \frac{h'(\eta)}{h(\eta)}.
\eeas
For the sake of convenience we have redefined
\beas
g(\eta)
&:=&
4 \pi \int_0^\eta \psi(\epsilon)(1+\eta-\epsilon)
\left[2(\eta-\epsilon) + (\eta-\epsilon)^2\right]^{1/2}\,d\epsilon,\\
h(\eta)
&:=&
\frac{4 \pi}{3} \int_0^\eta \psi(\epsilon)(1+\eta-\epsilon)
\left[2(\eta-\epsilon) + (\eta-\epsilon)^2\right]^{3/2}\,d\epsilon,
\eeas
and
\[
\psi (\epsilon) := \phi(E_0-\epsilon) = 
c \epsilon^k + O(\epsilon^{k+\delta})\ \mbox{as}\ \epsilon = E_0-E \to 0+.
\] 
In order to investigate the asymptotic behavior of $\beta$
we also need
\beas
g'(\eta)
&=&
4 \pi \int_0^\eta \psi(\epsilon)
\left[\left(2(\eta-\epsilon) + (\eta-\epsilon)^2\right)^{-1/2}
+ 2\,\left(2(\eta-\epsilon) + (\eta-\epsilon)^2\right)^{1/2}\right]
d\epsilon,\\
h'(\eta)
&=&
4 \pi \int_0^\eta \psi(\epsilon)
\left[\left(2(\eta-\epsilon) + (\eta-\epsilon)^2\right)^{1/2}
+ \frac{4}{3} \left(2(\eta-\epsilon) + (\eta-\epsilon)^2\right)^{3/2}\right]
d\epsilon.
\eeas
We need to substitute the 
asymptotic expansion of $\psi$ into $g$, $h$, and their derivatives; 
note that  $0<\eta(r)\to0$ as $r\to R-$.
The leading order terms always contain an integral of the form
\[
\int_0^{\eta} \epsilon^k (\eta-\epsilon)^m d\epsilon 
= \eta^{k+m+1} \int_0^1 s^k (1-s)^m\,ds =: \eta^{k+m+1} c_{k,m},\ m>-1,
\]
where the constants $c_{k,m}$ satisfy the relation
\[
\frac{c_{k,m-1}}{c_{k,m}}=\frac{k+m+1}{m},\ m>0.
\]
Hence for $\eta \to 0+$,
\beas
g(\eta)
&=&
4 \pi\, c\, 2^{1/2} c_{k,1/2}\; \eta^{k+3/2} + O(\eta^{k+3/2+\delta}),\\
h(\eta)
&=&
\frac{4 \pi}{3} c\, 2^{3/2} c_{k,3/2}\; \eta^{k+5/2} + O(\eta^{k+5/2+\delta}),\\
g'(\eta)
&=&
4 \pi\, c\, 2^{-1/2} c_{k,-1/2}\; \eta^{k+1/2} + O(\eta^{k+1/2+\delta}),\\
h'(\eta)
&=&
4 \pi\, c\, 2^{1/2} c_{k,1/2}\; \eta^{k+3/2} + O(\eta^{k+3/2+\delta}).
\eeas
If we substitute these expansions into the formula for $\alpha$
we find
\[
\alpha(r) = \frac{2}{3} \frac{c_{k,3/2}}{c_{k,1/2}} + O(\eta^\delta (r))
\to (k+3/2 +1)^{-1} > 0,\ r\to R-,
\]
Clearly, $\alpha (r) >0$ for $0<r<R$ and since $\eta(0) >0$,
\[
\lim_{r\to 0+}\alpha(r) = \frac{h(\eta(0))}{\eta(0)g(\eta(0))}>0
\]
so that
$\inf_{r\in]0,R[} \alpha(r) >0$ as desired. As to $\beta$,
\[
\beta(r) = \frac{c_{k,-1/2}}{c_{k,1/2}} -
\frac{3}{2} \frac{c_{k,1/2}}{c_{k,3/2}} + O(\eta^\delta (r))
\to k+\frac{1}{2},\ r\to R-,
\]
and by the assumption on $k$ this limit lies in the interval
$]0,2[$. We can therefore apply Lemma~\ref{lm:makino} to conclude
that $R<\infty$, and the proof of 
Proposition~\ref{finradss} is complete.
\prfe 

Although the class of steady states obtained by the approach above
is quite large,
the condition on $k$ excludes for example polytropes with
$k\geq 3/2$ and possible generalizations of those.
Hence we present a second approach the idea of which is the following.
If one introduces the speed of light $c$ as a parameter into the
relativistic Vlasov-Poisson system then solutions of this system
tend to solutions of the non-relativistic case as $c\to \infty$.
Exploiting this for Eqn.~(\ref{semilinp}) one can show
that the potential in the relativistic case has to cross
the cut-off energy for sufficiently large $c$, if the corresponding
potential in the non-relativistic case does. In this way
one can show that essentially every compactly supported
steady state of the non-relativistic Vlasov-Poisson system has its 
relativistic counterpart. In particular, this will be true
for the polytropic
ansatz with $0\leq k<7/2$, which extends the range found in \cite{Ba}.
The perturbation argument outlined above was used for the
Vlasov-Einstein system in \cite{RR1}.

If the speed of light is not normalized to unity 
in the relativistic Vlasov-Poisson system the only change
is that $\sqrt{1+|v|^2}$ must be replaced by $\sqrt{1+|v|^2/c^2}$.
In the sequel we use the parameter $\gamma :=1/c^2$. In order
to obtain the correct non-relativistic limit as $\gamma \to 0$
we adjust the ansatz (\ref{isotropic}):
\be \label{ansatzc}
f(x,v)= \phi\left(\frac{1}{\gamma} \sqrt{1+\gamma |v|^2} +
U(x) -\frac{1}{\gamma}\right).
\ee
This ansatz satisfies the Vlasov equation with potential
$U$ and parameter $\gamma$.
As before the spatial density $\rho$ becomes a functional
of the potential $U$,
\[
\rho(r) = g_\phi(\gamma,U(r)),
\]
where
\[
g_\phi(\gamma,u)
:=4 \pi \int^{\infty}_{u}\phi(E)\left(1+\gamma\,(E-u)\right)
\left[2 (E-u) + \gamma (E-u)^2\right]^{1/2}\,dE,\ u\in \R,
\]
and it remains to analyze the Poisson equation for $U$ which
takes the form
\be \label{semilinpc}
U'(r) = \frac{4 \pi}{r^2}\int_0^r g_\phi(\gamma,U(s))\, s^2\, ds,\ r>0.
\ee
For $\gamma =0$ this is precisely the
equation which arises if one looks for steady states
of the non-relativistic Vlasov-Poisson system via the
ansatz
\be \label{ansatznr}
f(x,v) = \phi\left(\frac{1}{2}|v|^2 + U(x)\right),
\ee
and an analysis of the limit $\gamma \to 0$ yields the following
result.
\begin{prop} \label{ctoinf}
Let $\phi \in L^\infty_\mathrm{loc} (\R)$, $\phi \geq 0$,
and $\phi(E)=0,\ E\geq E_0$. Let $(f_0,U_0)$ be a 
non-trivial, compactly supported steady
state of the Vlasov-Poisson system obtained by the ansatz 
(\ref{ansatznr}), i.e., $U_0\in C^1([0,\infty[)$ solves
Eqn.~(\ref{semilinpc}) with $\gamma =0$,
and $U_0(0) <E_0$ but $U_0(R)>E_0$ for some $R>0$.
Then for all $\gamma >0$ sufficiently small the solution
$U \in C^1([0,\infty[)$ of (\ref{semilinpc}) with 
$U(0)=U_0(0)$ satisfies $U(R)>E_0$, i.e., the resulting 
steady state of the relativistic Vlasov-Poisson system
with parameter $\gamma$ has compact support and finite mass.
\end{prop}
\prf
It is easy to see that 
there exists a constant $C>0$ such that
\[
|g_\phi(\gamma,u) - g(0,u_0)| \leq C \gamma + C |u-u_0|,\ 
u,u_0\geq U_0(0).
\]
In what follows, constants denoted by $C$
are positive, may change from line to line,
and never depend on $\gamma$, $U$, or $r$.
By monotonicity the solutions $U$ and $U_0$
under consideration take only values $u\geq U_0(0)$,
and hence
\[
|U'(r) - U_0'(r)| \leq C \gamma + C \int_0^r |U(s) -U_0(s)|\, ds,
\ 0\leq r \leq R.
\]
If we integrate this estimate, observe that $U(0)=U_0(0)$,
and apply Gronwall's lemma we find that
\[
|U(r) - U_0(r)| \leq C \gamma,\ 0\leq r \leq R.
\]
The latter holds in particular for $r=R$,
and since by assumption $U_0(R)>E_0$ the same
is true for $U$, provided $\gamma$ is small enough.
\prfe

\smallskip

\noindent
{\bf Remark.}
The requirement that $\gamma$ must be sufficiently small
in the result above actually changes the relativistic
Vlasov-Poisson system we started with. However, it is easy to
check that if $f$ is a steady state of the system
with some $\gamma >0$ then the rescaled function 
$f_\gamma (x,v):= \gamma^{-3/2} f(\gamma^{-1/2}x,\gamma^{-1/2}v)$
is a steady state of the original system with $\gamma=1$.
This scaling changes the cut-off energy, which however
has to be adjusted in order to satisfy the boundary condition
$U(\infty)=0$ anyway, and it changes the ansatz,
$f_\gamma (x,v) = \gamma^{-3/2} \phi((\sqrt{1+|v|^2} +U(x) -1)/\gamma)$,
but it does not change the principal way in which the
ansatz depends on the particle energy.

\smallskip

In the non-relativistic case it is well known
that the polytropic ansatz with $0\leq k <7/2$
leads to compactly supported steady states;
as a matter of fact the compact support is
lost for $k=7/2$, the so-called Plummer sphere,
and finite mass is lost for $k>7/2$. 
Generalizations of these polytropes were obtained by variational
techniques in \cite{GR3}, and by techniques which are 
closer to the ones used to prove Proposition~\ref{finradss}
in \cite{HRU}, so there are many examples which one can extend
to the relativistic case by Proposition~\ref{ctoinf}.

To conclude this section we 
emphasize that its purpose was only to demonstrate
that there are many examples of steady states which
satisfy the assumptions of our stability analysis.
Its purpose was not a systematic investigation of the possible
steady states of the relativistic Vlasov-Poisson system.

\bigskip

\noindent
{\bf Acknowledgment.} It is a pleasure for the authors
to thank Prof.~Y.~Guo for helpful discussions on the 
subject of this paper.

\end{document}